\documentclass[12pt]{article} 
\usepackage{graphicx}
\usepackage{epsfig}
\usepackage{amsbsy}

\begin{document}
\baselineskip 10mm

\centerline{\large \bf Anomalous Thermal Stability of Metastable C$_{20}$
Fullerene}

\vskip 6mm

\centerline{I. V. Davydov, A. I. Podlivaev, and L. A. Openov}

\vskip 4mm

\centerline{\it Moscow Engineering Physics Institute (State
University), 115409 Moscow, Russia}

\vskip 8mm

\centerline{\bf ABSTRACT}

The results of computer simulation of the dynamics of fullerene C$_{20}$
at different temperatures are presented. It is shown that, although it
is metastable, this isomer is very stable with respect to the transition
to a lower energy configuration and retains its chemical structure under
heating to very high temperatures, $T\approx 3000$ K. Its decay activation
energy is found to be $E_a\approx 7$ eV. Possible decay channels are
studied, and the height of the minimum potential barrier to decay is
determined to be $U=5.0$ eV. The results obtained make it possible
to understand the reasons for the anomalous stability of fullerene C$_{20}$
under normal conditions

\newpage

\centerline{\bf 1. INTRODUCTION}

\vskip 2mm

After the discovery of fullerene C$_{60}$ [1], the interest in
carbon clusters has greatly increased, due both to their unusual
physical and chemical properties and to their prospects for
practical applications [2]. In spite of numerous experimental and
theoretical studies, the mechanisms of carbon cluster formation
and some of their properties are still not clear.

The smallest of the experimentally observed "three-dimensional" carbon
clusters is the C$_{20}$ cluster [3], which is one of the fullerenes with a
sphere-like structure having carbon atoms located at their "surface" at the
vertices of pentagons or hexagons. In fullerene C$_{20}$, there are only
pentagons. A C$_{20}$ cluster can exist both in the form of a fullerene (a
cage) and in the form of a bowl, ring, chain, etc. (Fig. 1). The problem of
relative stability of these isomers remains to be fully resolved.
Experimental data are still incomplete and inconsistent, and theoretical
calculations performed using various methods give appreciably different
results [4-11]. Partly, this is due to the difficulty in finding the
correlation contribution to the total energy of the cluster and also to the
fact that the differences in the isomer energies are comparable to the errors
of calculation in the methods used.

Nevertheless, most authors who use the most exact modern computing
algorithms agree that, of all C$_{20}$ isomers, the bowl has the
minimum energy, whereas the cage is a metastable configuration [8,
11]. At the same time, there is no question that, experimentally
[3], both C$_{20}$ bowls and C$_{20}$ cages have been synthesized
[12, 13]. Thus, there is a question as to why fullerene C$_{20}$
retains an energetically unfavorable chemical structure under real
experimental conditions and does not pass to the lower energy
configuration.

In this study, the energy and structural characteristics of some
C$_{20}$ isomers are calculated using the tight-binding method
with a "transferable" potential of interatomic interaction. The
dynamics of fullerene C$_{20}$ is studied in detail at different
temperatures. It is shown that, although this three-dimensional
isomer is metastable, its lifetime under normal conditions is very
long because of a high potential barrier separating the metastable
state from the lower energy atomic configuration.

\vskip 6mm

\centerline{\bf II. METHODS OF CALCULATION}

\vskip 2mm

To calculate the energies of different configurations of the
C$_{20}$ cluster, we used the tight-binding method with a
transferable interatomic potential suggested for carbon compounds
in [14]. This method differs favorably from the majority of
empirical approaches and allows one to more correctly determine
the contribution of the electronic subsystem to the total energy.
Thus, four valence electrons of each carbon atom are taken into
account and the interatomic potential is actually an $N$-particle
potential, where $N$ is the number of atoms in the system.
Although the tight-binding method is not as exact as the {\it ab
initio} methods, it adequately describes both small carbon
clusters and macroscopic forms of carbon [7, 14] and, in addition,
strongly simplifies calculations even for rather large clusters.
In particular, in using the molecular dynamics method, this method
makes it possible to collect statistics sufficient for estimating
the decay activation energy and the lifetime of the metastable
state. Earlier, we applied this method in computer simulation of a
metastable C$_8$ cluster [15-18].

To find equilibrium and metastable configurations of a C$_{20}$ cluster, we
used the method of structural relaxation. First, an initial configuration of
atoms was chosen, which then relaxed to a state corresponding to a global or
local energy minimum under the action of intracluster interactions only.
At each time step of the relaxation, the velocities of all atoms were
decreased by (1 - 10) $\%$, which is equivalent physically to cooling of the
system. The time step was $t_0 = 2.72\cdot 10^{-16}$ s, which was
approximately equal to one percent of the vibration period for a C$_2$ dimer.

To determine the decay activation energy of the metastable configuration,
we used the method of molecular dynamics with a transferable tight-binding
potential (tight-binding molecular dynamics, TBMD [7]) and a time step $t_0$.
Calculations were performed at a fixed total energy, which corresponds to the
case of a heat-insulating system. The temperature $T$ of the cluster was
determined from the formula [15]
\begin{equation}
\frac{3}{2}k_BT=\langle E_{kin}\rangle~,
\label{Ekin}
\end{equation}
where $k_B$ is the Boltzmann constant and $\langle E_{kin}\rangle$ is the ion
kinetic energy per atom averaged over several vibration periods.

When calculating the forces ${\bf F}_i$ acting on the atoms ($i$ is the atom
number), we assumed the electron temperature $T_{el}$ to be equal to $T$ and
used the formula
\begin{equation}
{\bf F}_i=-2\sum_{n}\langle\Psi_n|\nabla_i\hat{H}|\Psi_n\rangle
f(\varepsilon_n)-\nabla_iU \label{Fi}
\end{equation}
which is a generalization of the Hellmann-Feynman formula to finite
temperatures [19, 20]. Here, $U$ is the classical component of the total
energy taking into account the repulsion of atoms at close distances;
$\hat{H}$ is the electron Hamiltonian in the tight-binding approximation
[14]; $|\Psi_n\rangle$ and $\varepsilon_n$ are the eigenstates and
eigenenergies of $\hat{H}$, respectively ($n = 1 - 80$); and
$f(\varepsilon_n$) is the Fermi-Dirac distribution function.
The chemical potential was determined at each step of molecular dynamics
simulation from the condition that the total number of valence electrons was
constant, $N_{el}=80$. To find the effect of the heating of the electron
subsystem on the cluster dynamics, we also performed calculations at
$T_{el}=0$.

The height of the potential barrier preventing the transition of the system
from the metastable configuration to a state with a lower energy was
calculated by the same method that we used previously in [16]. In this
method, calculation reduces to finding the saddle points of the potential
energy of the system considered as a function of the coordinates of all
atoms. These saddle points correspond to unstable equilibrium positions of
atoms in the cluster and possess the property that infinitesimal deviations
from the equilibrium positions result either in relaxation of the system to
the initial state or in a transition to a new configuration. To find saddle
points, the cluster is deformed continuously in the $3N$-dimensional space of
atomic coordinates along the direction of the vibration mode with a minimum
frequency so that the cluster energy monotonically increases with
deformation, while at the same time having local minima in all
others directions (orthogonal to the one chosen) [16].

\vskip 6mm

\centerline{\bf III. RESULTS}

\vskip 2mm

We calculated the structural and energy characteristics of four C$_{20}$
isomers: a cage, a bowl (Fig. 1), a ring, and a chain. For each isomer,
the binding energy $E_b$ was calculated from the formula
\begin{equation}
E_b=20E(C_1)-E(C_{20})~,
\label{Eb}
\end{equation}
where $E$(C$_{20}$) is the binding energy of a C$_{20}$ cluster
and $E$(C$_1$) is the energy of an isolated carbon atom. The
configuration with a maximum energy $E_b$ is stable (equilibrium),
since its total energy is minimum. The configurations with smaller
(but positive) values of $E_b$ are metastable; they correspond to
local minima of the total energy in the space of atomic
coordinates.

We obtained the following values of the binding energy $E_b$ per atom:
6.08, 6.14, 5.95, and 5.90 eV/atom for a cage, a bowl, a ring, and a chain,
respectively (Table 1). Our results indicate that the C$_{20}$ cage is
metastable, which is in agreement with Monte Carlo calculations [8, 11].
The bond lengths between the nearest neighbors in the C$_{20}$ fullerene are
listed in Table 2; these results are in good agreement with the results
obtained by other authors.

Heating can transform a metastable isomer to another
configuration. The characteristic time of such transformation (the
lifetime $\tau$) depends on temperature and the height of the
energy barrier separating these configurations. Following general
arguments [15], we can see that the cluster decay probability $W$
per unit time is given by the statistics formula
\begin{equation}
W=W_0\exp(-E_a/k_BT)~ ,
\label{W}
\end{equation}
where the factor $W_0$ has dimensions of inverse time (s$^{-1}$) and $E_a$
is the activation energy for cluster decay. This energy is close to
the height of the minimum energy barrier separating the metastable state from
the equilibrium state or from another metastable state but can differ
from it due to the presence of several different decay paths. The cluster
lifetime can be defined as [15]
\begin{equation}
\tau=1/W=\tau_0\exp(E_a/k_BT)~ ,
\label{tau}
\end{equation}
where $\tau_0=1/W_0$. It is convenient to pass over from the cluster lifetime
to the critical number of steps of molecular-dynamics simulation $N_c$
corresponding to cluster decay:
\begin{equation}
N_c=N_0\exp(E_a/k_BT)~ ,
\label{Nc}
\end{equation}
where $N_0=\tau_0/t_0$.

We performed molecular-dynamics simulation of the "life" of the C$_{20}$
fullerene at different initial temperatures $T$ of the ionic subsystem;
in this way, we directly determined $N_c$ as a function of $T$. Different
values of $T$ corresponded to different sets of initial velocities of the
cluster atoms ${\bf V}_{i0}$, which were chosen randomly each time (but
subjected to the condition $\sum_{i}{\bf V}_{i0}=0$).

The results obtained are shown in Fig. 2. Since the nature of the decay of a
metastable state is probabilistic, the quantity $N_c$ at a given temperature
$T$ is not determined uniquely. Nevertheless, it is seen from Fig. 2 that, in
first approximation, the results of simulation are described by Eq. (6),
according to which the dependence of $\ln(N_c)$ on $1/T$ is linear. The slope
of this line is the activation energy for the cluster decay and is found to
be $E_a=8\pm 1$ eV at the temperature of the electronic subsystem $T_{el}=T$
and $E_a=7 \pm 1$ eV at $T=0$.

Figure 3 shows the results of calculating the "potential
landscape" for a C$_{20}$ cluster in the vicinity of the
metastable cage configuration (point {\it 1} in Fig. 3). Saddle
point {\it 2} is the nearest to the cage state and corresponds to
a configuration in which two C-C bonds begin to break and two
adjoining octagons form (Fig. 4). The energy of this configuration
is 4 eV higher than that of a cage. Analysis of the data of
molecular dynamics simulation shows that, though this
configuration actually appears from time to time during thermal
vibrations, the cluster does not decay. The reason for this
behavior is that the energy of the metastable state at point {\it
3}, which is the nearest to saddle point {\it 2} (Fig. 3), is only
0.1 eV lower than the energy at the saddle point (visually, atomic
configurations {\it 2} and {\it 3} are practically the same; each
of them has two octagons). Therefore, after arriving at a new
metastable state, the system does not stay there but, due to the
thermal motion of atoms, returns (again via saddle point {\it 2})
to the vicinity of the initial metastable state at point {\it 1}.

Metastable saddle point {\it 4}, next to metastable state {\it 3}
(Fig. 3), corresponds to a configuration in which two C-C bonds
are broken and the breaking of the third bond begins, resulting in
the formation of a cluster of three adjoining octagons on the
"lateral surface" (Fig. 5). The energy of this configuration is
4.8 eV higher than that of a cage. Metastable state {\it 5}, which
is the nearest to saddle point {\it 4}, also has three octagons
(and three broken C-C bonds). Molecular-dynamics simulation shows
that, after passing to metastable state {\it 5}, the system can
either return to the vicinity of metastable state {\it 1} via
saddle points {\it 4} and {\it 2} or pass to metastable state {\it
7} via saddle point {\it 6} (Fig. 3). In atomic configuration {\it
6}, three C-C bonds are broken and the breaking of another bond
begins; because of this, four octagons are formed on the lateral
surface of the cluster (Fig. 6). In metastable configuration {\it
7}, there are also four octagons. As a rule, the system does not
return from this configuration to the original state {\it 1} (we
observed such a return only once). Thus, the difference between
the energies of configurations {\it 6} and {\it 1} is the height
$U=5.0$ eV of the minimum potential barrier preventing the cage
decay.

Having passed over this barrier and appeared in metastable state {\it 7},
the system, in the overwhelming majority of cases, very rapidly passes via
saddle point {\it 8} to metastable state {\it 9} (Fig. 3), where it resides
for a time corresponding to $10^3 \div 10^4$ steps of molecular dynamics
simulation. Configuration {\it 9} has the form of a star and is shown in
Fig. 7. In this symmetric configuration, there are five octagons on the
lateral surface of the cluster. The decay of the star leads to the formation
of different quasi-two-dimensional or quasi-one-dimensional configurations
(Fig. 8) and occurs through transitions via different saddle points (only one
of them is shown in Fig. 3). No transition occurs to the equilibrium bowl
configuration.

\vskip 6mm

\centerline{\bf IV. DISCUSSION}

\vskip 2mm

It should be noted that the problem of the choice of the temperature of the
electron subsystem $T_{el}$ in simulating the dynamics of an electron-ion
system is not at all trivial. It is shown in [19] that the use of Eq. (2) in
integrating the classical equations of motion for ions corresponds to the
conservation of the so-called Mermin free energy [22] $\Omega=E-T_{el}S$,
where $E$ is the total internal energy of the system and $S$ is the
electronic entropy. Generally, the quantity $T_{el}$ does not necessarily
coincide with the average ionic temperature $T$ [19]. Calculations of the
dynamics of different fullerenes at high temperatures performed in [20]
showed, in particular, that the stability of the cluster at $T_{el}=T$
appears to be somewhat lower than at $T_{el}=0$ but that there are no basic
qualitative distinctions between these two cases. According to [20], the
fragmentation temperature $T_{fr}$ of fullerene C$_{20}$, defined as the
temperature above which the metastable configuration decays, is
$T_{fr}\approx 3600$ K and 4000 K at $T_{el}=T$ and $T_{el}=0$, respectively.

According to our data, the activation energy $E_a$ for the decay
of fullerene C$_{20}$ is the same at $T_{el}=T$ and $T_{el}=0$ to
within the limits of error (see Section 3), although the average
value of $E_a$ at $T_{el}=T$ is somewhat greater than that at
$T_{el}=0$. Thus, the cluster dynamics weakly depends on the
temperature of the electronic subsystem. Partly, this is due to a
rather large gap (0.4 eV) between the energies of the lowest
unoccupied and the highest occupied molecular orbital (HOMO-LUMO
gap). Here, we should emphasize that the "physical time" during
which we controlled the dynamics of the C$_{20}$ cluster for each
set of initial velocities (i.e., at each initial temperature)
exceeded 1.5 ns, which is two orders of magnitude greater than the
corresponding time (about 10 ps) in [20, 23], where simulations of
the thermal stability of fullerenes were performed. Due to this,
we could find the temperature dependence of $N_c$ (i.e., the
cluster lifetime) over a fairly large temperature range and
estimate the activation energy $E_a$. Obviously, the lifetime of a
metastable state is dependent on temperature. Therefore, it makes
no sense to determine (as in [20]) "the temperature of cluster
fragmentation" regardless of the time in which this fragmentation
occurs.

Using the calculated energy $E_a$, the lifetime of a C$_{20}$ cage
at room temperature, $\tau$(300 K), is found from Eq.(5) to be
very large (practically, infinite). This result allows us to
understand the reason for the success of experiments [3] on the
synthesis of metastable C$_{20}$ fullerenes. We note that the
height of the minimum potential barrier to the decay of a C$_{20}$
cage ($U=5.0$ eV) is somewhat lower than the decay activation
energy $E_a=(6\div 9)$ eV determined directly from the molecular
dynamics simulation data. This is partly due to the fact that the
cluster can decay in different ways by passing through potential
barriers of different heights, including those that are higher
than the lowest barrier.

In [24], calculations of the temperature dependence of the relative
root-mean-square fluctuation of bond lengths $\delta$ led to the conclusion
that the C$_{20}$ cluster melts at a temperature $T_m\approx  1900$ K.
According to our calculations, $\delta$ monotonically increases with
temperature without exhibiting any features at $T < 3000$ K. Moreover, an
abrupt cooling of the cluster at any instant prior to its decay results
either in its transition to one of the intermediate metastable states (states
{\it 3}, {\it 5} in Fig. 3) or in its return to the original metastable state
(state {\it 1}). Thus, the problem of melting of the C$_{20}$ fullerene at a
certain temperature $T_m$ requires further study.

Let us now discuss in more detail the character of transition of the C$_{20}$
fullerene to other states. Above all, we note that we never observed a
transition to the equilibrium bowl configuration with a lower total energy
(higher binding energy). As a rule, the decay of the C$_{20}$ fullerene
begins with a transition to the metastable star configuration via a sequence
of several saddle points and intermediate short-lived metastable states
(Fig. 3). For a star (Fig. 7), the binding energy $E_b=5.91$ eV/atom is lower
than that for a cage. A transition from the cage to the star is accompanied
by a decrease in the cluster temperature by (500 - 800) K. Over the course of
time, the star passes to different (as a rule, quasi-two-dimensional or
quasi-one-dimensional) configurations with a lower binding energy and the
cluster temperature decreases from (3000 - 4000) K to (1000 - 1500) K. These
configurations are very different in structure from both the cage and the
bowl (one of such configurations is shown in Fig. 8). Thus, the ability of
carbon structures to form numerous intermediate metastable (but fairly
stable) states prevents the transition of the metastable cage to the stable
bowl configuration.

\vskip 6mm

\centerline{\bf V. CONCLUSIONS}

The main result of this study is that the metastable C$_{20}$ fullerene
(cage) was demonstrated to have very high thermal stability with respect to
transition to an equilibrium state with a lower total energy. The reason for
this stability is the high potential barrier, which prevents the decay of the
C$_{20}$ cage and the corresponding high decay activation energy. For this
reason, the lifetime of the C$_{20}$ cage is very long even at room
temperature. Therefore, once created at a certain stage of synthesis, a
C$_{20}$ cage retains its chemical structure.

Although all this applies to an isolated C$_{20}$ cage, by analogy to a
C$_{60}$ cluster, we may hope that a C$_{20}$ fullerene-based cluster
material (fullerite) exists. In any case, the preliminary data are rather
encouraging [21, 25, 26]. Final solution of this problem requires further
experimental and theoretical studies, one stimulus for which is the
conjecture that the C$_{20}$ fullerite (if synthesized) could be a
high-temperature superconductor [27].

\vskip 6mm

\centerline{\bf ACKNOWLEDGMENTS}

\vskip 2mm

This study was supported by the CRDF, project "Scientific and Educational
Center for Basic Research of Matter in Extremal States".

\newpage

Table 1. Binding energies $E_b$ (eV/atom) for some isomers of C$_{20}$
clusters calculated using different methods: tight-binding method (TB),
Hartree-Fock method (HF), density functional method in the local density
approximation (LDA), density functional method with gradient corrections
(GCA), and Tersoff-Brenner empirical method (Tersoff)

\vskip 6mm

\vbox{\offinterlineskip
\hrule
\halign{&\vrule#&
\strut\quad\hfil#\quad\cr
height2pt&\omit&&\omit&\cr
&C$_{20}$ isomer\hfil&&TB [6]&&HF [8]&&LDA [8]&&GCA [9]
&&Tersoff [21]&&This work&\cr
\noalign{\hrule}
height2pt&\omit&\cr
&Cage&&6.08&&4.01&&7.95&&6.36&&6.36&&6.08&\cr
&Bowl&&-&&4.15&&7.87&&-&&6.19&&6.14&\cr
&Ring&&6.01&&4.23&&7.77&&6.45&&6.11&&5.95&\cr
&Chain&&6.05&&-&&-&&6.35&&-&&5.90&\cr
height2pt\cr}
\hrule}

\vskip 46mm

Table 2. Bond lengths in the C$_{20}$ fullerene calculated using
different methods: Hartree-Fock method (HF), method of modified neglect
of differential overlap (MNDO), and Tersoff-Brenner empirical method
(Tersoff)

\vskip 6mm

\vbox{\offinterlineskip
\hrule
\halign{&\vrule#&
\strut\quad\hfil#\quad\cr
height2pt&\omit&&\omit&\cr
&Method\hfil&&$l_{min}$, A&&$l_{max}$, A&\cr
\noalign{\hrule}
height2pt&\omit&\cr
&HF [4]&&1.42&&1.47&\cr
&MNDO [5]&&1.41&&1.52&\cr
&Tersoff [21]&&1.44&&1.53&\cr
&This work&&1.44&&1.52&\cr
height2pt\cr}
\hrule}

\newpage

\includegraphics[width=0.4\textwidth]{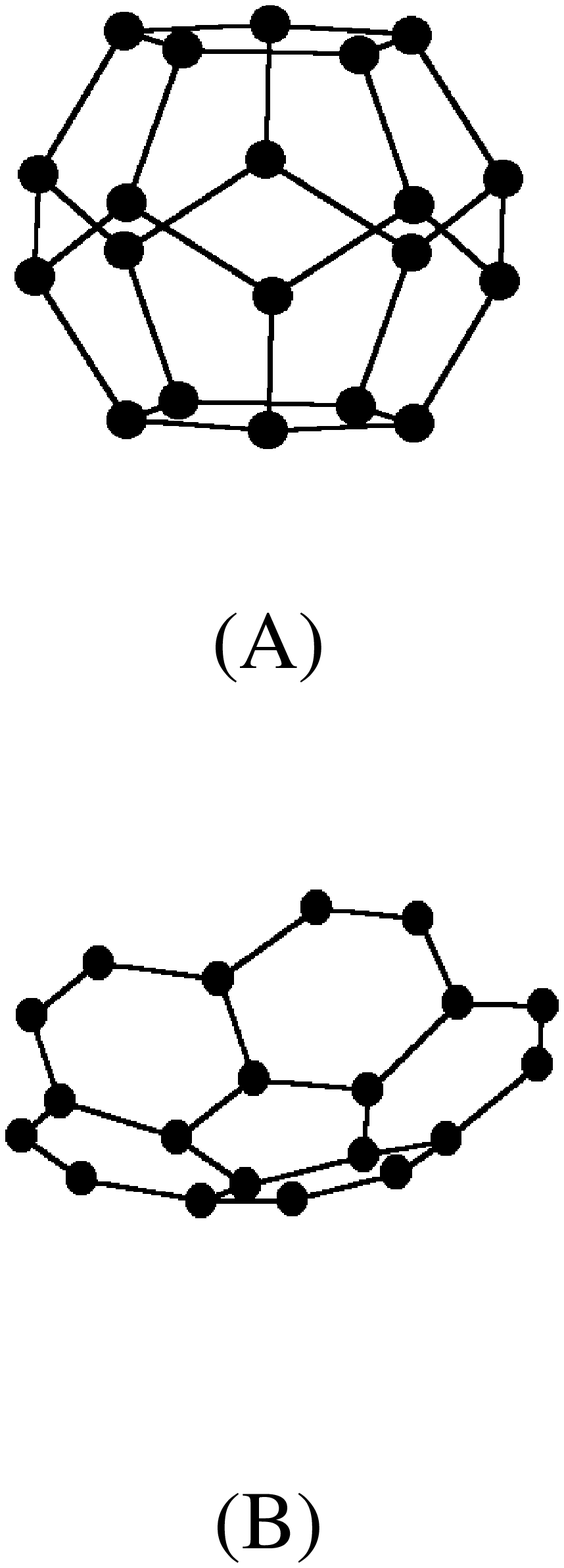}

\vskip 6mm

Fig. 1. Isomers of a C$_{20}$ cluster. (a) Fullerene (cage) and (b) bowl.

\newpage

\includegraphics[width=0.5\textwidth]{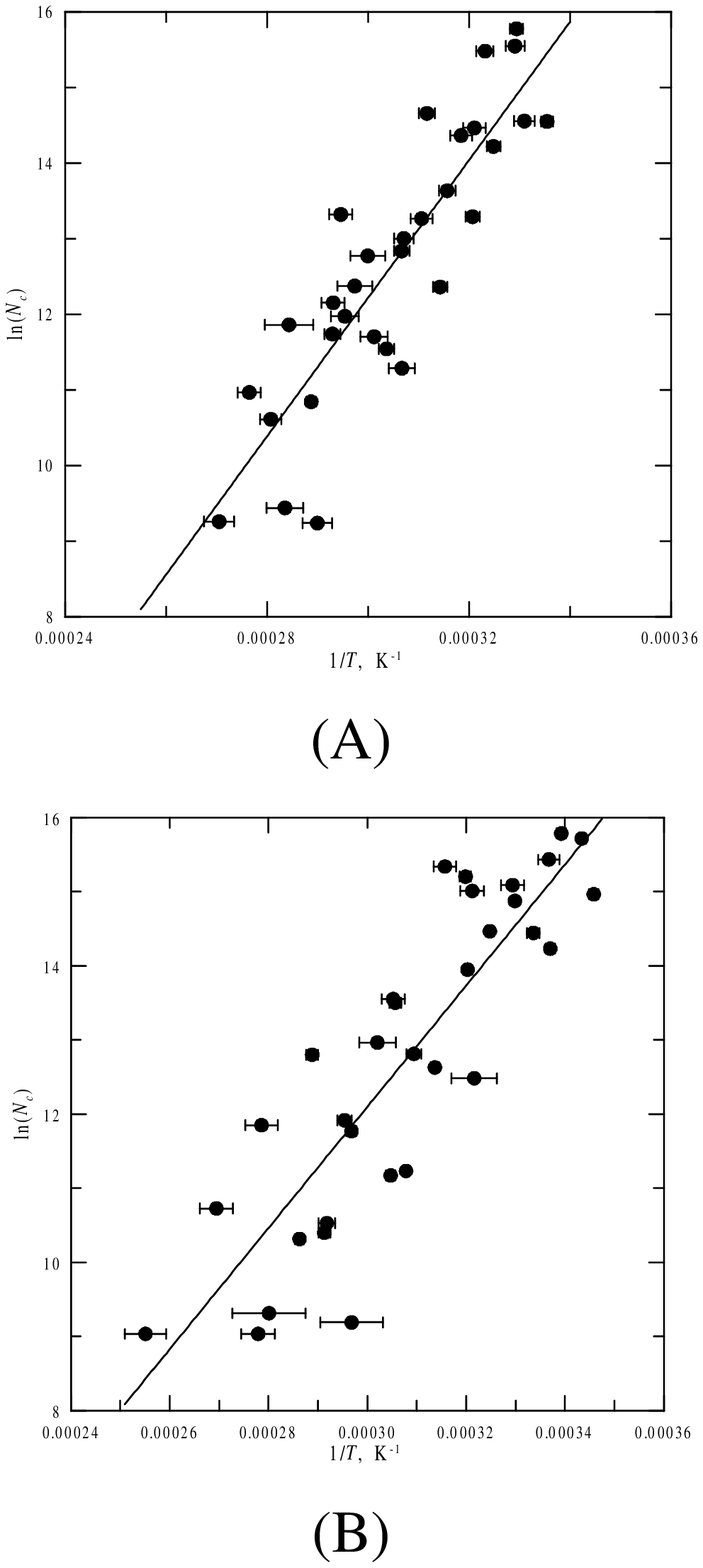}

\vskip 6mm

Fig. 2. Logarithm of the critical number of steps of molecular dynamics
simulation $N_c$ for the onset of the decay of the C$_{20}$ fullerene as a
function of the temperature $T$ of the ionic subsystem for electron
temperature (a) $T_{el}=T$ and (b) $T_{el}=0$. Circles are the results of
calculation, and the solid line is a linear least squares fit.

\newpage

\includegraphics[width=0.7\textwidth]{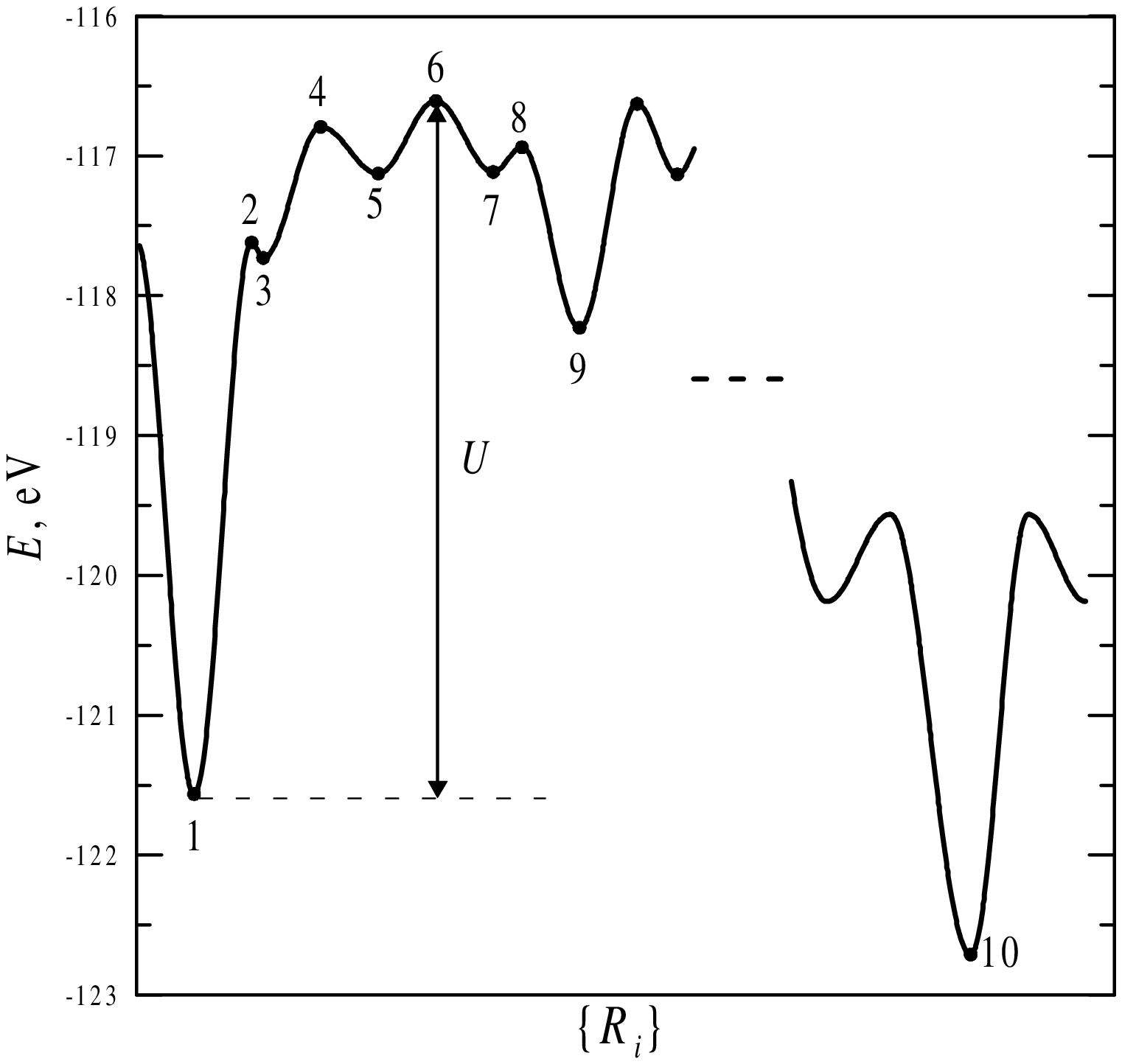}

\vskip 6mm

Fig. 3. Dependence of the total energy $E$ of a C$_{20}$ cluster on
"the generalized coordinate" in the $3N$-dimensional space of atomic
coordinates ${\bf R}_i$ in the vicinity of the metastable cage configuration
(schematic). The energies are measured from the energy of 60 isolated carbon
atoms. The numerals correspond to the following configurations: ({\it 1})
fullerene (cage), $E=-121.56$ eV (Fig. 1a); ({\it 2}) saddle point,
$E=-117.62$ eV (Fig. 4); ({\it 3}) metastable state, $E=-117.73$ eV;
({\it 4}) saddle point, $E=-116.79$ eV (Fig. 5); ({\it 5}) metastable state,
$E=-117.13$ eV; ({\it 6}) saddle point determining the height of the minimum
potential barrier ($U=5.0$ eV) to the decay of the cage, $E=-116.61$ eV
(Fig. 6); ({\it 7}) metastable state, $E=-117.12$ eV; ({\it 8}) saddle point,
$E=-116.94$ eV; ({\it 9}) metastable star state, $E=-118.23$ eV (Fig. 7);
and ({\it 10}) equilibrium bowl configuration, $E=-122.71$ eV (Fig. 1b).

\newpage

\includegraphics[width=\hsize]{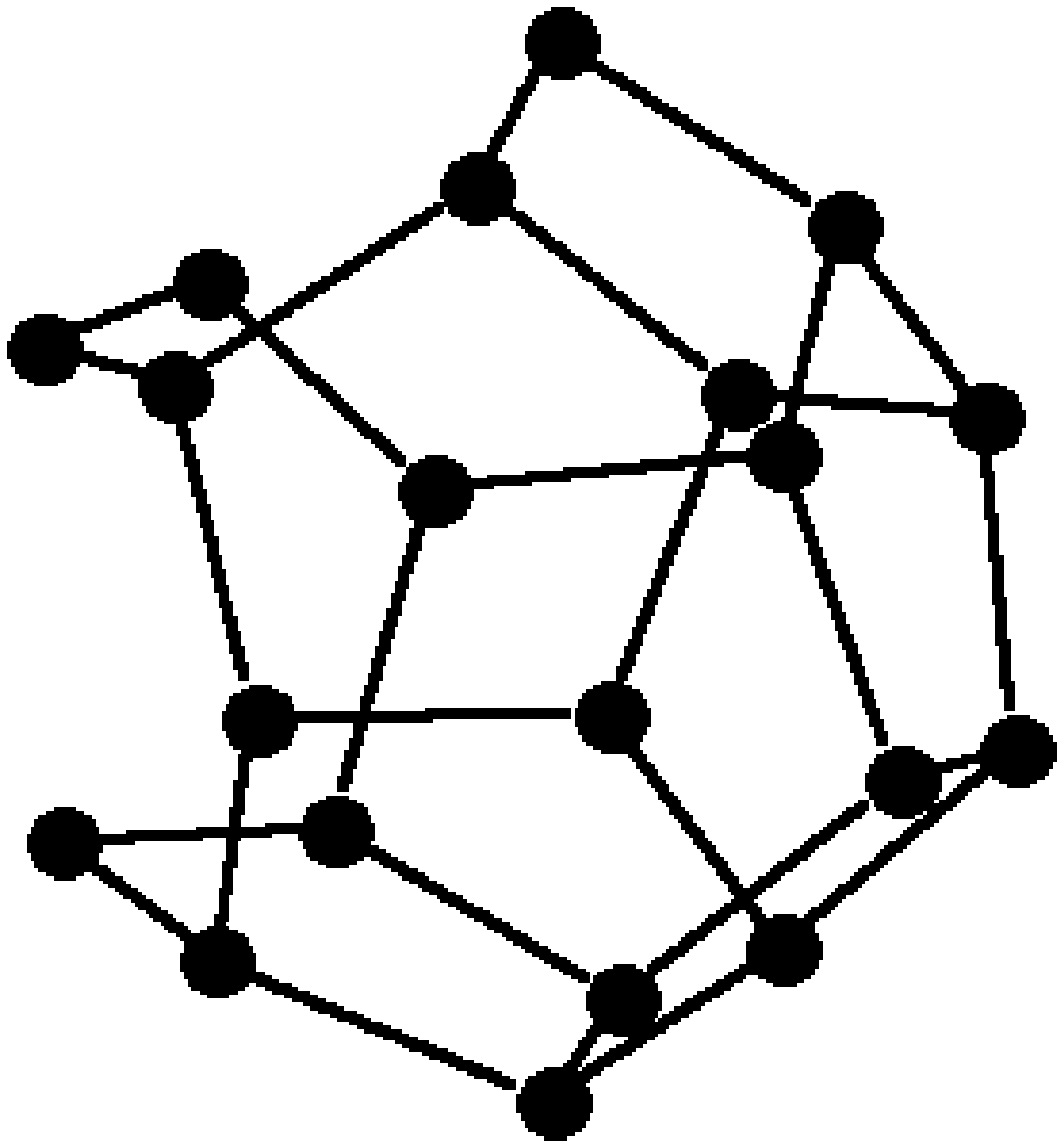}

\vskip 6mm

Fig. 4. Atomic configuration corresponding to saddle point {\it 2} in Fig. 3.

\vskip 6mm

\newpage

\includegraphics[width=\hsize]{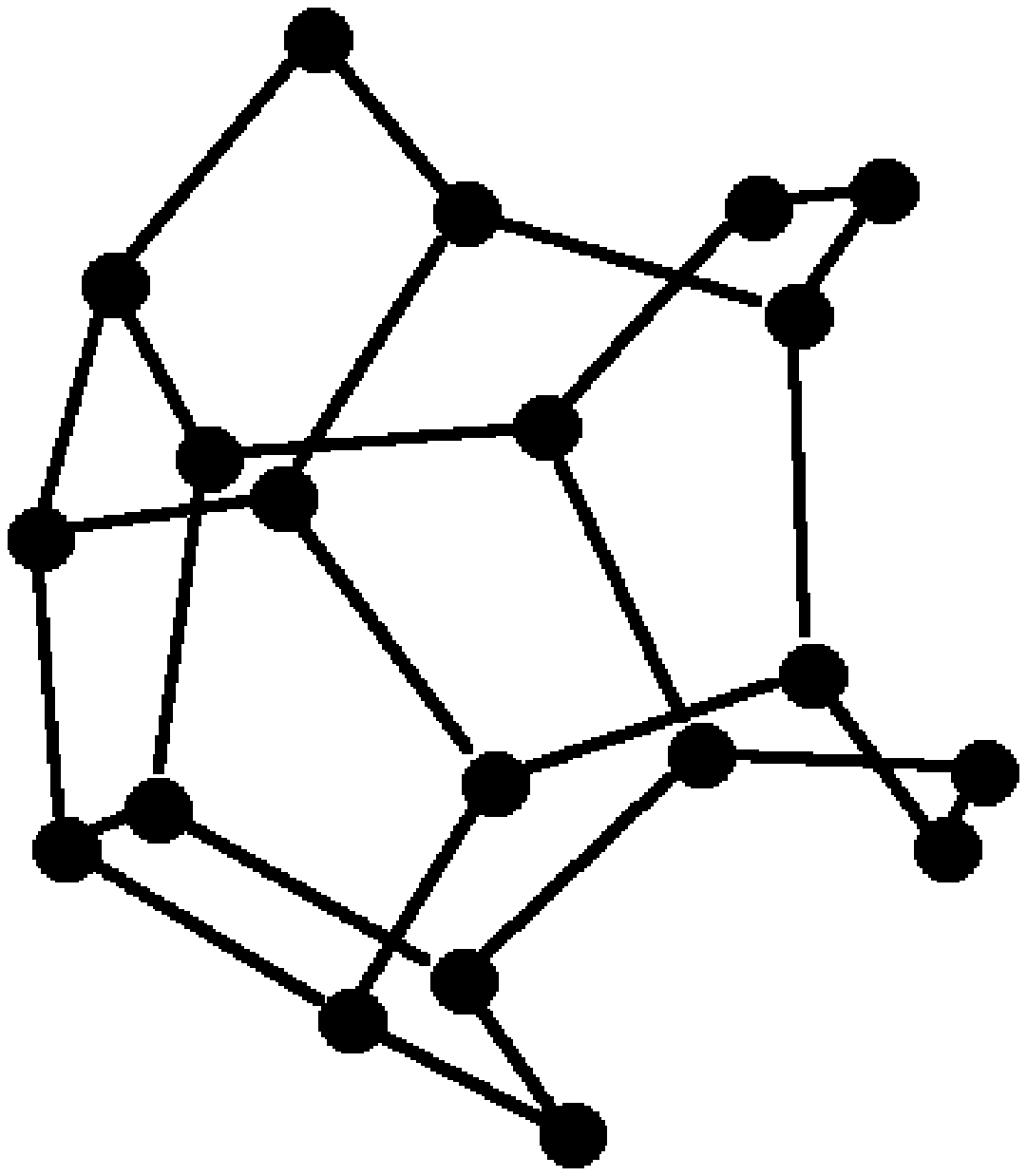}

\vskip 6mm

Fig. 5. Atomic configuration corresponding to saddle point {\it 4} in Fig. 3.

\newpage

\includegraphics[width=\hsize]{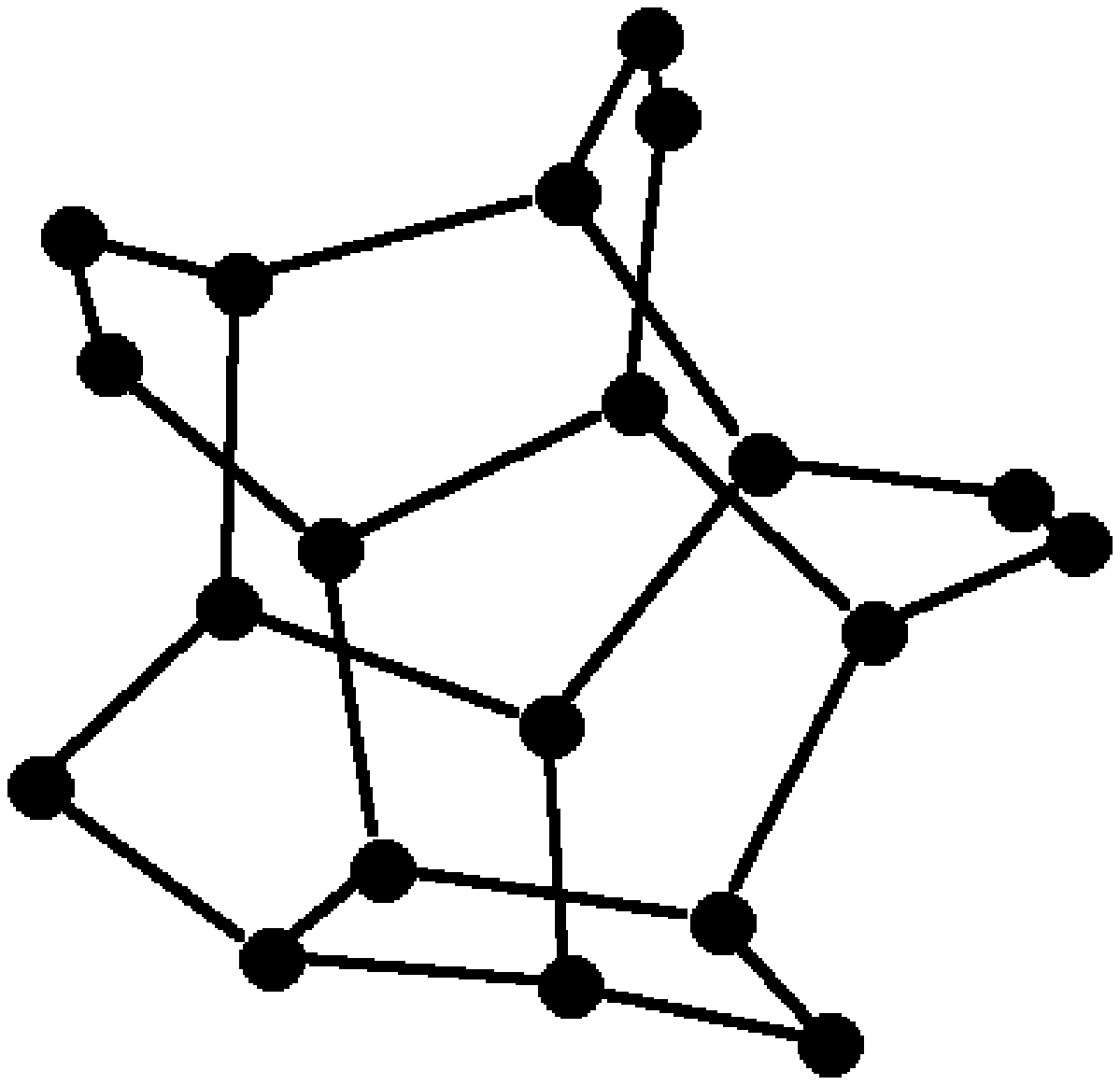}

\vskip 6mm

Fig. 6. Atomic configuration corresponding to saddle point {\it 6} in Fig. 3.
This configuration determines the height of the minimum potential barrier
($U=5.0$ eV) to the decay of the cage.

\newpage

\includegraphics[width=\hsize]{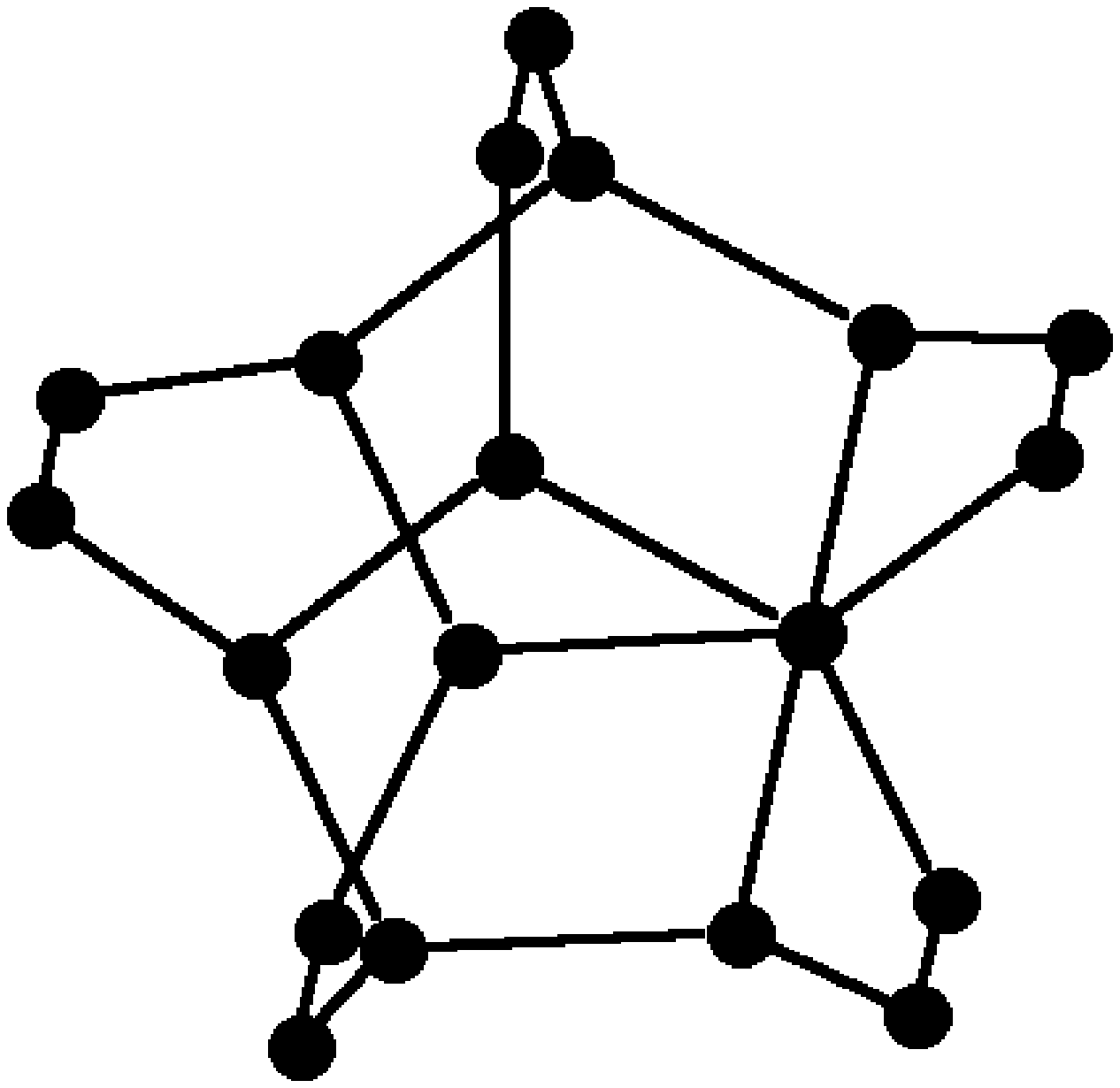}

\vskip 6mm

Fig. 7. Atomic configuration "star" corresponding to metastable state {\it 9}
in Fig. 3.

\newpage

\includegraphics[width=\hsize]{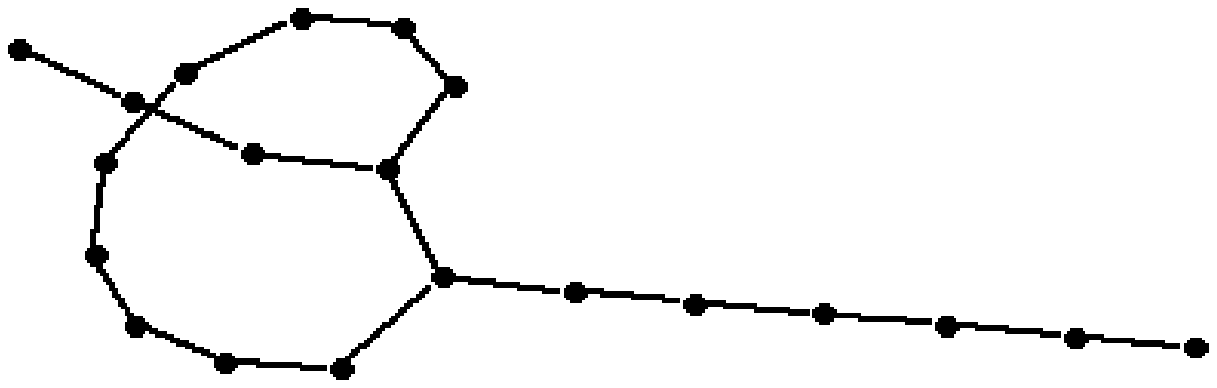}

\vskip 6mm

Fig. 8. One of the atomic configurations formed after the decay of the star.


\vskip 6mm

\begin{thebibliography}{5}

\bibitem{1} H. W. Kroto, J. R. Heath, S. C. O'Brien, R. F. Curl,
and R. E. Smalley, Nature {\bf 318}, 162 (1985).

\bibitem{2} A. V. Eletskii and B. M. Smirnov,
Usp. Fiz. Nauk {\bf 165}, 977 (1995) [Phys. Usp. {\bf 38}, 935 (1995)].

\bibitem{3} H. Prinzbach, A. Weller, P. Landenberger, F. Wahl, J. Worth,
L. T. Scott, M. Gelmont, D. Olevano, and B. von Issendorff,
Nature {\bf 407}, 60 (2000).

\bibitem{4} V. Parasuk and A. Alml\"{o}f,
Chem. Phys. Lett. {\bf 184}, 187 (1991).

\bibitem{5} D. Bakowies and W. Thiel,
J. Am. Chem. Soc. {\bf 113}, 3704 (1991).

\bibitem{6} D. Tom\'{a}nek and M. A. Schluter,
Phys. Rev. Lett. {\bf 67}, 2331 (1991).

\bibitem{7} C. H. Xu, C. Z. Wang, C. T. Chan, and K. M. Ho,
Phys. Rev. B {\bf 47}, 9878 (1993).

\bibitem{8} J. C. Grossman, L. Mitas, and K. Raghavachari,
Phys. Rev. Lett. {\bf 75}, 3870 (1995).

\bibitem{9} R. O. Jones and G. Seifert,
Phys. Rev. Lett. {\bf 79}, 443 (1997).

\bibitem{10} R. O. Jones, J. Chem. Phys. {\bf 110}, 5189 (1999).

\bibitem{11} S. Sokolova, A. L\"{u}chow, and J. B. Anderson,
Chem. Phys. Lett. {\bf 323}, 229 (2000).

\bibitem{12} M. Saito and Y. Miyamoto,
Phys. Rev. Lett. {\bf 87}, 035503 (2001).

\bibitem{13} J. Lu, S. Re, Y. Choe, S. Nagase, Y. Zhou, R. Han, L. Peng,
X. Zhang, and X. Zhao, Phys. Rev. B {\bf 67}, 125415 (2003).

\bibitem{14} C. H. Xu, C. Z. Wang, C. T. Chan, and K. M. Ho,
J. Phys.: Condens. Matter {\bf 4}, 6047 (1992).

\bibitem{15} L. A. Openov and V. F. Elesin, Pis'ma Zh. Eksp. Teor. Fiz.
{\bf 68}, 695 (1998) [JETP Lett. {\bf 68}, 726 (1998)];
physics/9811023.

\bibitem{16} V. F. Elesin, A. I. Podlivaev, and L. A. Openov,
Phys. Low-Dim. Struct. {\bf 11/12}, 91 (2000); physics/0104058.

\bibitem{17} L. A. Openov and V. F. Elesin, Mol. Mater. {\bf 13}, 391 (2000).

\bibitem{18} N. N. Degtyarenko, V. F. Elesin, N. E. L'vov, L. A. Openov,
and A. I. Podlivaev, Fiz. Tverd. Tela (St. Petersburg) {\bf 45}, 954 (2003)
[Phys. Solid State {\bf 45}, 1002 (2003)].

\bibitem{19} R. M. Wentzcovitch, J. L. Martins, and P. B. Allen,
Phys. Rev. B {\bf 45}, 11 372 (1992).

\bibitem{20} B. L. Zhang, C. Z. Wang, C. T. Chan, and K. M. Ho,
Phys. Rev. B {\bf 48}, 11 381 (1993).

\bibitem{21} A. J. Du, Z. Y. Pan, Y. K. Ho, Z. Huang, and Z. X. Zhang,
Phys. Rev. B {\bf 66}, 035405 (2002).

\bibitem{22} N. D. Mermin, Phys. Rev. [Sect. A] {\bf 137}, 1441 (1965).

\bibitem{23} S. G. Kim and D. Tom\'{a}nek,
Phys. Rev. Lett.{\bf 72}, 2418 (1994).

\bibitem{24} X. Z. Ke, Z. Y. Zhu, F. S. Zhang, F. Wang, and Z. X. Wang,
Chem. Phys. Lett. {\bf 313}, 40 (1999).

\bibitem{25} V. Paillard, P. M\'{e}linon, V. Dupuis, A. Perez, J. P. Perez,
G. Guiraud, J. Fornazero, and G. Panczer,
Phys. Rev. B {\bf 49}, 11433 (1994).

\bibitem{26} S. Okada, Y. Miyamoto, and M. Saito,
Phys. Rev. B {\bf 64}, 245405 (2001).

\bibitem{27} I. Spagnolatti, M. Bernasconi, and G. Benedek,
Europhys. Lett. {\bf 59}, 572 (2002).

\end{thebibliography}
\end{document}